\documentclass[11pt]{article}
\usepackage{amsmath,amssymb,latexsym,amsfonts}
\usepackage{times}
\usepackage{amsthm}
\usepackage[dvips]{graphics}
\begin{document}
\bibliographystyle{plain}

\newtheorem{theorem}{Theorem}[section]
\newtheorem{corollary}{Corollary}[section]
\newtheorem{lemma}{Lemma}[section]
\newtheorem{claim}{Claim}[section]
\newtheorem{exercise}{Exercise}
\newtheorem{definition}{Definition}[section]
\newtheorem{remark}{Remark}[section]
\newtheorem{conjecture}{Conjecture}[section]
\newtheorem{note}{Note}[section]
\newtheorem{example}{Example}[section]
\newtheorem{fact}{Fact}[section]
\newtheorem{proposition}{Proposition}[section]

\font\boldsets=msbm10
\def\F{{\hbox{\boldsets  F}}}
\def\N{{\hbox{\boldsets  N}}}
\def\R{{\hbox{\boldsets  R}}}
\def\Z{{\hbox{\boldsets  Z}}}
\def\A{{\hbox{\boldsets  A}}}
\def\G{{\hbox{\boldsets  G}}}
\def\blackslug{\hbox{\hskip 1pt \vrule width 4pt height 8pt depth 1.5pt
\hskip 1pt}}

\newcommand {\half}{\frac 1 2}
\newcommand{\ignore}[1]{}
\newcommand {\tr}{\mbox{tr}}

\newcommand{\bbox}{\vrule height7pt width4pt depth1pt}
\def\QED{\quad\blackslug\lower 8.5pt\null\par}
\def\proof{\par\penalty-1000\vskip .5 pt\noindent{\bf Proof\/: }}
\newcommand{\prb}{\begin{problem}}
\newcommand{\eprb}{\bbox\end{problem}}
\newcommand {\upint}[1]   {\lceil {#1} \rceil}



\title
{Are stable instances easy?
}
\author{
Yonatan Bilu
\thanks{
This research is supported by grants from the binational Science
Foundation Israel-US and the Israel Science Foundation. }
\\
Mobileye Vision Technologies Ltd.\\
12 Hartom Street, PO Box 45157\\
Jerusalem, 91450 Israel.\\
yonatan.bilu@mobileye.com
\and Nathan Linial
\footnotemark[1]
\\
Institute of Computer Science\\
Hebrew University\\
Jerusalem 91904, Israel.\\
nati@cs.huji.ac.il
}
\maketitle

\begin{abstract}
We introduce the notion of a stable instance for a discrete
optimization problem, and argue that in many practical situations
only sufficiently stable instances are of interest. The question
then arises whether stable instances of NP--hard problems are
easier to solve. In particular, whether there exist algorithms
that solve correctly and in polynomial time all sufficiently
stable instances of some NP--hard problem. The paper focuses on
the Max--Cut problem, for which we show that this is indeed the
case.
\end{abstract}

\section{Introduction}
Computational complexity theory as we know it today is concerned
mostly with worst-case analysis of computational problems. For
example, we say that a problem is NP-hard if the existence of an
algorithm that correctly decides {\em every} instance
of the problem implies that
SAT can be decided in a polynomially equivalent time complexity.
However, the study of decision and optimization problems is
motivated not merely by theoretical considerations. Much of our
interest in such problems arises because they formalize certain
real-world tasks. From this perspective, we are not interested in
{\em all} problem instances, but only in those which can actually
occur in reality.

This is often the case with clustering problems, which are
ubiquitous in most fields of engineering, experimental and applied
science. Any concrete formulation of the clustering problem is likely to be
NP-hard. However this does not preclude the possibility that the
problem can be solved efficiently in practice. In fact, in numerous
application areas, large-scale clustering problems are solved on a regular
basis. As mentioned above, we are only interested in instances
where the data is actually made up of fairly well-defined clusters
- the instances where solving the problem is interesting from the
practical perspective.

Put differently, the usual way for proving that clustering is
NP-hard is by a reduction to, say, SAT. This reduction entails the
construction of instances for the clustering problem, such that the
existence of an algorithm that can solve all of them efficiently
implies the existence of an algorithm that efficiently solves SAT.
However, it may well be the case that all these instances are
clearly artificial, and solving them is of no practical interest.

As a concrete example, consider the problem of clustering protein
sequences into families. Out of the enormous space of all possible
sequences, only a tiny fraction is encountered in nature, and it
is only about these (or slight modifications thereof) that we
actually care.

Our case in point is the Max-Cut problem, which can be thought of
as a clustering into two clusters. It is well known that this
problem is NP-complete, and so it is believed that there is no
algorithm that solves it correctly on {\em all} graphs, in
polynomial time. In this work we strive to identify properties of
instances of the Max-Cut problem (i.e., of weighted graphs), which
capture the notion that the input has a well-defined structure
w.r.t Max-Cut (i.e., the maximal cut ``stands out'' among all
possible cuts). Our goal is to show that Max-Cut can be solved
efficiently on inputs that have such properties.

Consideration of a similar spirit have led to
the development of {\em Smoothed Analysis} initiated in \cite{SpTeng},
(see \cite{Vershy} for some of the exciting developments in that area.
The similarity has two main facets: (i) Both lines of research
attempt to investigate the computational
complexity of problems from a non-worst-case perspective,
(ii) Both are investigations of the {\em geometry} of the
instance space of the problem under consideration. The
goal being to discover interesting parts of this space in which
the instances have complexity lower than the worst case.
Viewed from this geometric perspective, the set-up that we study
here is very different than what is done in the theory of
smoothed analysis. There one shows that the hard instances
form a discrete and isolated subset of the input space.
Consequently, for every instance of the problem, a small random
perturbation is very likely to have low computational complexity.
In the problems that we study here the situation is radically
different. The ``interesting'' instances ({\em stable} instances
as we shall call them) are very rare. Indeed, it is not hard to show that
under reasonable models of random instances the probability that
a random instance be stable is zero,
or at least tends to zero as the problem size grows.
What we wish to accomplish is to efficiently solve {\em all} instances
within this subspace. We claim that this tiny set is interesting
because it includes all realistic clustering problems.

The notion of {\em stability} is central to our work. This is a
concrete way to formalize the notion that
the only instances of interest are
those for which small perturbation in the data (which may reflect
e.g. some measurement errors) do not change the optimal partition
of the graph.

\begin{definition}
Let $W$ be an $n \times n$ symmetric, non-negative matrix. A {\em
$\gamma$-perturbation} of $W$, for $\gamma \geq 1$, is an $n
\times n$ matrix $W'$ such that $\forall i,j=1,\ldots,n$,
$W_{i,j} \leq W'_{i,j} \leq \gamma \cdot W_{i,j}$.\\
Let $(S,[n] \backslash S)$ be a maximal cut of $W$, i.e. a
partition that maximizes $\sum_{i\in S, j \notin S} W_{i,j}$. The
instance $W$ (of the Max-Cut problem) is said to be
$\gamma$-stable, if for every $\gamma$-perturbation $W'$ of $W$,
$(S,[n] \backslash S)$ is the unique maximal cut of $W'$.
\end{definition}

However this definition is, perhaps, not sufficient.
Consider two bipartite graphs which are
joined togther by a single edge. The resulting graph is $\gamma$-stable
for all $\gamma$, but the alignment of the two bipratite graphs
with respect to one another completely depends on the adjoining edge.
Hence, to better capture our intution of what it means for a solution
to be stable, it is reasonable to demand that in addition to stability
the graph contains no small cuts. We show that the combination of both these 
properties indeed allows solving Max-Cut efficiently (Example~\ref{example:cheeger}).

In section \ref{sec:comb} we present an algorithm that solves
correctly and in polynomial time $\gamma$-stable instances of
Max-Cut: (i) On simple graphs of minimal degree $\delta$, when
$\gamma > \frac {2n} \delta$, and (ii) On weighted graphs of
maximal degree $\Delta$ when $\gamma > \sqrt{\Delta n}$. In
section \ref{sec:spect} we explore several spectral conditions
which make Max-Cut amenable on stable instances. This involves
analyzing the {\em spectral partitioning} heuristic for Max-Cut.
In particular, we show that Max-Cut can be solved efficiently
on (locally) stable expander graphs, and on graphs where the
solution is sufficiently distinct from all other cuts.
We conclude
by deducing an improved approximation bound for the
Goemans-Williamson algorithm on stable instances, and by showing
that Max-Cut is easy in a certain random model for such instances.

Finally, we should mention that this is just a first step. In
particular, it is of great interest to study more permissive
notions of stability where a small perturbation can slightly modify
the optimal solution. There are also other natural ways to capture the
concept of stability. Similar considerations can be applied to
many other optimization problems. Some of these possibilities
are briefly discussed below, but these questions are mostly
left for future investigations.

\section{Preliminaries}
\subsection{Notation}
Throughout the paper we denote the vertex set of the graph $G$
under discussion by $[n]$. A vector $v \in \R^n$ induces the
partition of $[n]$ into the sets. $( \{i : v_i > 0\}, \{i : v_i
\leq 0\})$. Viewed as a partition of $G$'s vertex set, we call it
the {\em cut induced by $v$} in $G$.

The {\em indicator vector} of a partition $(S, \bar{S})$ of $[n]$
(or a cut in $G$), is the vector $v \in \{-1,1\}^n$, with $v_i = 1$ iff $i \in S$.\\
For a weighted graph $G$, we denote the indicator vector of its
maximal cut by $mc^*$. We generally assume that this cut is
unique, otherwise $mc^*$ is
an indicator of some maximal cut.

For a subset $A \subset [n]$, we denote $\bar{A} = [n] \backslash A$.

For two disjoint subsets of vertices in the graph, $A,B$, we
denote by $E(A,B)$ the set of edges going between them, and 
$w(A,B) =  \sum_{(i,j) \in E(A,B)} W_{i,j}$.
With a slight abuse of notation, we denote $w(i) =
\sum_j W_{i,j}$.
Finally, for a set
of edges $F \subset E$, denote $w(F) = \sum_{(i,j) \in F}
W_{i,j}$. 

We switch freely between talking about the graph and about
its associated weight matrix. Given a symmetric nonnegative 
 $n \times n$ matrix $W$ with zero trace (as input to the
max cut problem), we define its {\em support} as
a graph $G=(V,E)$ with vertex set $V=[n]$ where $(i,j) \in E$ iff
$w_{ij} >0$.

\subsection{Properties and equivalent definitions}
A useful way to think of $\gamma$-stability is as a game between two (computationally
unbounded) players, Measure and Noise: Given a graph $G$, Measure chooses
a cut $(S, \bar{S})$. Noise then multiplies weights of his choice
by factors between $1$ and $\gamma$, obtaining a graph $G'$ (over the same vertex
and edge sets, but with possibly different weights). He then chooses a different cut,
$(T, \bar{T})$. Noise wins if in $G'$  $w(T, \bar{T}) > w(S, \bar{S})$.
Otherwise, Measure wins. 
A graph is $\gamma$-stable if Measure has a winning strategy.

Observe that the players' strategy is clear: Measure chooses the maximal cut,
and Noise , w.l.o.g., multiplies by $\gamma$
the weights of the edges in
$E(T, \bar{T}) \backslash E(S, \bar{S})$. Multiplying weights of other
edges either does not change  $w(T, \bar{T}) - w(S, \bar{S})$, or
decreases it. Hence, we arrive at an equivalent definition for  $\gamma$-stability
is:

\begin{proposition}\label{def:gamelike}
Let $\gamma \geq 1$. A graph $G$ graph with  
maximal cut $(S, \bar{S})$ is $\gamma$-stable
(w.r.t. Max-Cut) if for
every vertex set
$T \neq S, \bar{S}$, 
\[w(E(S, \bar{S}) \backslash E(T, \bar{T}))> \gamma \cdot w(E(T, \bar{T}) \backslash E(S, \bar{S})).\]
\end{proposition}

This view of stability suggests how $\gamma$-stable graphs
can be generated: Let $G'$ be
a $\gamma'$-stable graph. Multiplying the weights of all the edges in the maximum cut by
$\frac {\gamma} {\gamma'}$ yields a $\gamma$-stable graph $G$. Moreover,
it is not hard to see that all $\gamma$-stable graphs can be obtained this way.
In other words, in the following random process every $\gamma$-stable graph on $n$
vertices has a positive probability: Generate a random graph on $n$ vertices, say
according to $G(n,p)$ (for some $ p \neq 0, 1$); Find the maximal cut and its
stability $\gamma'$. Multiply all cut edges by $\frac {\gamma} {\gamma'}$.
Note, however, that in this naive model the maximal cut can be easily identified by simply
examining edge weights - those of weight  $\frac {\gamma} {\gamma'}$ are the cut edges.

One pleasing aspect of $\gamma$-stability is that it is {\em oblivious to scale} -
multiplying all weights in a graph by a constant factor does not change its
stability. This can be readily seen from Proposition \ref{def:gamelike}.
It may seem natural to define
$\gamma$ two-way stability as robustness to perturbation by a multiplicative factor
between $\frac 1 {\gamma}$ and ${\gamma}$ (so called, two-way perturbation).
But obliviousness to scale easily implies that a graph is
$\gamma$ two-way stable iff it is $\gamma^2$-stable.

It is also natural to consider a solution as ``interesting'' if
it stands out among all the alternatives. Let $(S, \bar{S})$ be a maximal cut
in a graph $G$, and consider an alternative cut, $(T, \bar{T})$. Consider the set
$E(S, \bar{S}) \Delta E(T, \bar{T})$ of those edges on which the two cuts ``disagree''.
We seek to measure the difference between the cuts $(S, \bar{S})$
and $(T, \bar{T})$ relative to the size of
$w(E(S, \bar{S}) \Delta E(T, \bar{T}))$. So say that  $(S, \bar{S})$ is
{\em $\alpha$ edge distinct} (with $\alpha > 0$), if for any $T \subset V$,
\[w(S, \bar{S}) - w(T, \bar{T}) > \alpha \cdot w(E(S, \bar{S}) \Delta E(T, \bar{T})).\]

Now, denote $W_T =  w(E(T, \bar{T}) \backslash E(S, \bar{S}))$ and
$W_S = w(E(S, \bar{S}) \backslash E(T, \bar{T}))$. If $G$ is $\alpha$ edge distinct
then
\[W_S - W_T = w(S, \bar{S}) - w(T, \bar{T}) > \alpha \cdot w(E(S, \bar{S}) \Delta E(T, \bar{T})) =
\alpha \cdot (W_S + W_T).\]
Hence, $W_S \geq \frac {\alpha} {1 - \alpha} W_T$, and by Proposition \ref{def:gamelike}
$G$ is $ \frac {1 + \alpha} {1 - \alpha}$-stable.
Similarly, if $G$ is $\gamma$-stable, then it is $\frac {\gamma - 1} {\gamma+1}$ edge distinct.

\subsection{Variations on a theme}

We shall also be interested in a weaker version of stability, which proves
useful for some of the results in sequel:

\begin{definition}
Let $W$ be an instance of the Max-Cut problem and let $(S,\bar{S})$
be its optimal partition. We say that $W$ is
{\em $\gamma$-locally stable} if for all $v \in S$
\[\gamma \cdot \sum_{u \in S}W_{u,v} < \sum_{u \in \bar{S}}W_{u,v},\]
and for all $v \in \bar{S}$
\[\gamma \cdot \sum_{u \in \bar{S}}W_{u,v} < \sum_{u \in S}W_{u,v},\]
\end{definition}

Observe that every $\gamma$-stable graph is also $\gamma$-locally
stable - this follows from Definition \ref{def:gamelike},
with $T$ being a single vertex.

It is essentially known that Max-Cut is NP-hard even when
restricted to $\gamma$-locally stable instances
(for $\gamma$ at most exponential in
the size of the input) \cite{Papadi}
\footnote{The NP-completeness of Max-Cut can be shown by a reduction from
$3$-Not-all-equal SAT: Construct a graph over the formula's literals, and for
every $3$-clause define three edges (a triangle) connecting the clause's literals.
It is not hard to see that the formula is satisfyable iff the graph's Max-Cut's value 
is twice the number of clauses. It is also not hard to see that if this is indeed the case,
the cut is $2$-locally stable. Furthermore, by adding edges between a literal and its negation,
the structure of the Max-Cut does not change, and local stability increases.
}.
In fact, one can impose
local stability, without altering the overall stability: Let $G$
be a graph with weighted adjacency matrix $W$. Let $G^{\times}$ be
a graph on $V \times \{0,1\}$, with weighted adjacency matrix:
\begin{eqnarray*}
G^{\times} =
\left(
\begin{array}{cc}
W & \tau \cdot w(i) \cdot I \\
\tau \cdot w(i) \cdot I & W
\end{array}
\right)
\end{eqnarray*}
(for some $\tau \geq 1$.)\\
It is not hard to see that the maximal cut in $G^{\times}$ consists of
two copies of that in $G$. Specifically, $(S,\bar{S})$ is a maximal cut
in $G$ iff
$(S \times\{0\} \cup \bar{S}  \times\{1\},
S \times\{1\} \cup \bar{S}  \times\{0\})$ is a maximal cut
in $G^{\times}$.\\
It is also not hard to see that $G$ is $\gamma$-stable, iff
$G^{\times}$ is, and that $G^{\times}$ is at least $2 \tau$-locally stable.

The definition of stability via edge distinctness formalizes the notion that in
instances of interest, the Max-Cut should be distinctly better
than all other cuts. Clearly, cuts which differ only slightly from
the maximum one in structure can only differ slightly in value, so
the difference in value should be quantified in terms of of the
distance between the cuts.
\begin{definition}
Let $(S, \bar{S})$ be
a cut in a (weighted) graph $G=(V,E)$ and $k > 0$. We say that this cut
is {\em $k$-distinct} if for any cut $(T, \bar{T})$,
\[
w(e(S, \bar{S})) - w(e(T, \bar{T}) \geq k\min\{|S \Delta T|, |S \Delta \bar{T}|\}.
\]
We say that a graph is {\em $(k,\gamma)$-distinct} (w.r.t.
Max-Cut) if its maximal cut is $k$-distinct and $\gamma$-locally
stable.
\end{definition}

In example \ref{example:distinct} we show that Max-Cut can be
solved on $(k,\gamma)$-distinct instance when $k$ and $\gamma$ are
sufficiently large.

\section{Combinatorial approach}\label{sec:comb}
One approach in solving a Max-Cut problem is to identify a pair of 
vertices which must to be on 
the same side of the optional cut 
(e.g. in a simple graph, two vertices with the same neighborhood).
Two such vertices can be safely merged into a single vertex - keeping multiple edges. 
If this can be repeated until a bipartite graph is obtained,
then the problem is solved.

Observe that if $G$ is a $\gamma$-stable graph, and $i$, $j$ are two vertices
on the same side of the maximal cut, then the graph $G'$, obtained from $G$ by
merging $i$ and $j$ into a single vertex $i'$, is  $\gamma$-stable as well. 
Indeed, any $\gamma$-perturbation of $G'$ induces a $\gamma$-perturbation of $G$
over the same edges. If as a result of this perturbation the maximal cut changes in $G'$,
then this new cut is also maximal in the similarly perturbed $G$, since it contains
the same edges (in contradiction with $G$ being $\gamma$-stable).

This observation implicitly guides the first algorithm presented below. In it we 
identify pairs of vertices which are on opposite sides of the maximal cut. By continuing 
to do so, we grow bigger and bigger connected bipartite subgraphs, until they all connect.
In the second algorithm we explicitly merge together vertices on the same side as long
as we know how to, and then, once we have a much smaller graph, use the first algorithm.

\subsection{An efficient algorithm for $n$-stable instances}\label{sec-greedy}
We start by describing an algorithm, that solves the Max-Cut
problem on (weighted) graphs of maximal degree $\Delta$ which are
$\sqrt{\Delta n}$-stable. The idea is to iteratively identify sets
of edges which belong to the maximal cut. When they form a
connected spanning bipartite graph, the maximal cut is found.

\framebox{
\begin{minipage}{5in}
\vspace*{10pt}
\footnotesize
{\bf FindMaxCut($G$)} \hspace*{10pt} ($G$ is a weighted graph)
\begin{enumerate}
\item Initialize $T = (V(G),\emptyset)$.
Throughout the algorithm $T$ will be a bipartite subgraph of $G$.
\item While $T$ is not connected, do:
\begin{enumerate}
\item Let $C_1,\ldots,C_t$ be the connected components of $T$. Each of them
is a bipartite graph, with vertex bipartition $V(C_i) =
(L_i,R_i)$.
\item Let $C_{i^*}$ be a component with the least number of vertices.
For each $j=1,\ldots,t$, $j \neq i^*$, let $E_j^0 = E(L_{i^*},L_j)
\cup E(R_{i^*},R_j)$ and $E_j^1 = E(L_{i^*},R_j) \cup
E(R_{i^*},L_j)$. Let $j^*$ and $c^*$ be such that the weight of
$E_{j^*}^{c^*}$ is the largest among all
$E_{j}^{c}$.\label{step-astrx}
\item Add the edges of $E_{j^*}^{c^*}$ to $T$
\end{enumerate}
\item Output the cut defined by the two sides of $T$.
\end{enumerate}
\end{minipage}
}

\begin{theorem}\label{MC-greedy}
There is an algorithm that solves correctly and in polynomial time
every instance of weighted Max-Cut that is $\gamma$-stable for
every  $\gamma > \sqrt{\Delta n}$. Here an instance is an
$n$-vertex graph of maximal degree $\Delta$.
\end{theorem}
\proof We will show that the above algorithm is well defined, and
outputs the correct solution on $\sqrt{n \Delta}$-stable instances
of Max-Cut. Let $(S,\bar{S})$ be the maximal cut. We maintain that
throughout the algorithm, $S$ {\em separates} each connected
component $C_i = (L_i, R_i)$. Namely, either
$L_i \subset S, \ R_i \subset V \backslash S$ or $R_i \subset S, \ L_i \subset V \backslash S$.\\
This clearly holds at the outset. If it holds at termination, the
algorithm works correctly. So consider the first iteration when
this does not hold. Let $C_{i^*}$ be a smallest connected
component at this stage, and denote $k = |C_{i^*}|$. Up to this
point our assumption holds, so say $L_{i^*} \subset S$ and
$R_{i^*} \cap S = \emptyset$. Let $j^*$ and $c^*$ be those chosen
as in step \ref{step-astrx}. Since this is the point where the
algorithm errs, $E_{j^*}^{c^*}$ is added to
$T$, yet $E_{j^*}^{c^*} \cap E(S,\bar{S}) = \emptyset$.\\
Now consider the $\gamma$-perturbation of the graph obtained by multiplying
the edges in $E_{j^*}^{c^*}$ by $\gamma$. If the original graph is
$\gamma$-stable, the maximal cut of the perturbed graph is  $(S,\bar{S})$
as well. Consider the cut obtained by flipping the sides of $L_{i^*}$ and
$R_{i^*}$. That is, denote $Z = S \backslash
L_{i^*} \cup R_{i^*}$, and consider
the cut $(Z,\bar{Z})$.\\
The cut $(Z,\bar{Z})$ contains the edges $E_{j^*}^{c^*}$, which
$(S,\bar{S})$ does not. For each $j \neq j^*$, let $c_j$ be such that
$E_{j}^{c_j}$ is in the cut $(S,\bar{S})$ (we'll be interested only in
non-empty subsets).
In the extreme case, all these edges are not in the cut $(Z,\bar{Z})$.
Observe that all other edges in $E(S,\bar{S})$ are also in $E(Z,\bar{Z})$.\\
Define $J = \{ j \neq i: E_{j}^{c_j} \neq \emptyset \}$.
Since the weight of $(Z,\bar{Z})$, even in the perturbed graph, is smaller
than that of $(S,\bar{S})$, we have that:
\[\gamma \cdot w(E_{j^*}^{c^*}) <
\sum_{j \in J} w(E_{j}^{c_j}).\]
(The l.h.s. is a lower bound on what we gain when we switch from $S$ to $Z$, and
the r.h.s. is an upper bound on the loss.)
Recall that  $E_{j^*}^{c^*}$
was chosen to be the set of edges with the largest total weight. Hence,
$\sum_{j \in J} w(E_{j}^{c_j}) \leq |J| w(E_{j^*}^{c^*})$,
and so $\gamma < |J|$.
Clearly, $|J| \leq \min \{\frac n k, k \Delta\}$, and so:
\begin{eqnarray*}
\gamma^2 < \frac n k k \Delta = n \Delta.
\end{eqnarray*}
This is a contradiction to the assumption that the input is
$\sqrt{n \Delta}$-stable.
\QED

Note that we
have actually proven that the algorithm works as long as it can find
a connected component $C_{i^*}$, such that
$|\{ j : E_{j}^{c} \neq \emptyset \}| < \gamma$, for $c = 0,1$.

The concept of stability clearly applies to other
combinatorial optimization problems. Similarly, the algorithm
above can be adjusted to solve highly stable instances of other
problems. For example, a similar algorithm finds the optimal
solution to (weighted) $\sqrt{n \Delta}$-stable instances of the
Multi-way Cut problem, and $\Delta$-stable instances of the Vertex
Cover problem (where again $n$ is the number of vertices in the
graph, and $\Delta$ the maximal degree).

\subsection{An efficient algorithm for simple graphs of high minimal degree}\label{subsec-simple}
A complementary approach is useful when the graph is unweighted,
and of high minimal degree. Suppose a $\gamma$-stable graph, for
some big (but bounded) $\gamma$ has minimal degree $n/2$. Then by
local stability each side in the maximal cut must be of size
nearly $n/2$, and the neighborhoods of any two vertices on the
same side have most of their vertices in common. Thus we can
easily cluster together the vertices into the two sides of the
maximal cut. Even when the minimal degree is lower, we can use the
same scheme to obtain several clusters of vertices which are
certain to be on the same side, and then use the algorithm from
the previous subsection to find the maximal cut.

\begin{theorem}
There is an algorithm that solves correctly and in polynomial time
every instance of unweighted Max-Cut that is $\gamma$-stable for
every  $\gamma \ge \frac {2n} \delta$. Here an instance is an
$n$-vertex graph $G=(V,E)$ of minimal degree $\delta$. Furthermore, if
$\delta = \Omega(\frac n {\log n})$, then $\gamma$-local stability suffices.
\end{theorem}

It clearly suffices to consider $\gamma = \frac {2n} \delta$.
Let $N_i \subset V$ be the neighbor set of vertex $i$ and $d_i = |N_i|$.
Define $H$ to be a graph on $V$ with $i$, $j$ adjacent if $|N_i \cap N_j| > \frac {\min\{d_i,d_j\}}
{\gamma+1}$. Since $G$ is in particular $\gamma$-locally stable, every
vertex $i$ has at most $\frac {d_i} {\gamma+1}$ of its neighbors
on its own side of the maximal cut. Hence, the vertices of
each connected component of $H$ must be on the same side
of the maximal cut.

Let $c$ be the number of connected components in $H$ and let $U
\subset V$ be a set of $c$ vertices, with exactly one vertex from
each of these connected components. Let the degrees of the
vertices in $U$ be $d_{i_1} \leq  d_{i_2} \leq \ldots \leq
d_{i_c}$. For any $u, v \in U$ we have that  $|N_u \cap N_v| \leq
\frac {\min\{d_u,d_v\}} {\gamma+1}$. We claim that $c < \gamma$.
If this is not the case, let us apply the inclusion-exclusion
formula and conclude:

\[
|\bigcup_1^{\gamma} N_i| \ge
\sum_{j=1}^{\gamma} (d_{i_j} - \sum_{k=1}^{j-1} \frac {d_{i_k}} {\gamma + 1}) =
\sum_{j=1}^{\gamma} d_{i_j} (1 -  \frac{1}{\gamma + 1}\sum_{k=1}^{j-1} \frac {d_{i_k}} {d_{i_j}}) \ge
\sum_{j=1}^{\gamma} d_{i_j} (1 -  \frac{j-1}{\gamma + 1})
\]
since, by assumption $d_{i_k} \le d_{i_j}$ for $k < j$. Also,
$d_{i_j} \ge \delta$ for all $j$, and clearly $|\bigcup_1^{\gamma}
N_i| < n$. Therefore,
\[
n > |\bigcup_1^{\gamma} N_i| \geq \delta \sum_{j=1}^{\gamma} (1 - \frac {j-1} {\gamma + 1}) =
\delta (\gamma  -  \frac {\gamma(\gamma-1)}{2(\gamma + 1)}) \ge \frac{\gamma \delta}{2}
\]
a contradiction which implies $c <  \gamma$.

Now consider the graph $G'$ obtained from $G$ by contracting all
vertices in each $C_i$ into a single vertex, keeping multiple
edges. By our previous observation, $G'$ has the same max-cut
value as $G$. Consequently, as discussed at the beginning of
this section, the graph $G'$ is
$\gamma$-stable. It follows that $G'$ is a weighted graph whose
stability exceeds its number of vertices. By
Theorem~\ref{MC-greedy}, the optimal cut in $G'$ (and hence in
$G$) can be found in polynomial time, as claimed.

It is also worth mentioning that if $\delta = \Omega(\frac n {\log
n})$ then $\gamma$ is $O(\log n)$, and we can find the maximal cut
in $G'$ by going over all cuts. Moreover, in this case it suffices
to assume that $G$ is $\gamma$-locally stable. \QED

\section{A spectral approach}\label{sec:spect}
\subsection{Definitions}
{\em Spectral partitioning} is a general name for a number of
heuristic methods for various graph partitioning problems which
are popular in several application areas. The
common theme is to consider an appropriate eigenvector of a
possibly weighted adjacency matrix of the graph in question, and
partition the vertices according to the corresponding entries. Why
is it at least conceivable that such an approach should yield a
good solution for Max-Cut? The Max-Cut problem can can clearly be
formulated as:
\[\min_{y \in \{-1,1\}^n} \sum_{(i,j) \in E}W_{i,j}y_i y_j.\]
The Goemans-Williamson algorithm \cite{GoWi} works by solving an
SDP relaxation of the problem. In other words, where as above
we multiply the matrix $W$ by a rank $1$ PSD matrix, in the SDP
relaxation, we multiply it be a PSD matrix of rank (at most)
$n$. Let us consider instead the
relaxation of the condition $y \in \{-1,1\}^n$, to $y \in \R^n,
||y||^2 = n$. The resulting problem is well-known: By the
variational characterization of eigenvalues, this relaxation
amounts to finding the eigenvector corresponding to the least
eigenvalue of $W$. Let $u$ be such a vector. This suggests a {\em
spectral partitioning of $W$} that is the partition of $[n]$ induced by $u$.\\
We also consider what we call {\em extended spectral
partitioning}: Let $D$ be a diagonal matrix. Think of $W+D$ as the
weighted adjacency matrix of a graph, with loops added. Such loops
do not change the weight of any cut, so that regardless of what
$D$ we choose, a cut is maximal in $W$ iff it is maximal in $W+D$.
Furthermore, it is not hard to see that $W$ is $\gamma$-stable,
iff $W+D$ is. Our approach is to first find a ``good'' $D$, and
then take the spectral partitioning of $W+D$ as the maximal cut.
These observations suggest the following question: Is it true that
for every $\gamma$-stable instance $W$ with $\gamma$ large enough
there exists a diagonal $D$ for which extended spectral
partitioning solves Max-Cut? If so, can such a $D$ be found
efficiently? Below we present certain sufficient conditions for
these statements.

\subsection{Spectral partitions of stable instances}

The input to the max cut problem is a symmetric nonnegative $n
\times n$ matrix $W$ with zero trace. The {\em support} of $W$ is
a graph $G=(V,E)$ with vertex set $V=[n]$ where $(i,j) \in E$ iff
$w_{ij} >0$.

\begin{lemma}\label{lem-spec-part}
Let $W$ be a $\gamma$-stable instance of Max-Cut with support
$G=(V,E)$. Let $D$ be a diagonal matrix, and $u$ an eigenvector
corresponding to the least eigenvalue of $W+D$. If $\gamma \geq
\frac {\max_{(i,j) \in E} |u_i u_j|} {\min_{(i,j) \in E} |u_i
u_j|}$, then the spectral partitioning induced by $W+D$ yields the
maximal cut.
\end{lemma}
\proof As noted above, for any diagonal matrix $D$, the problems
of finding a maximal cut $W+D$ and in $W$ are equivalent.
Normalize $u$ so that ${\min_{(i,j)\in E} |u_i\cdot u_j| = 1}$. (If
$u$ has any $0$ coordinates, the
statement of the lemma is meaningless). Let $D'$ be the diagonal
matrix $D'_{i,i} = D_{i,i} \cdot u_i^2$. Let $W'$ be the matrix
$W'_{i,j} = W_{i,j} \cdot |u_i u_j|$. Observe that $W'$ is a
$\gamma$-perturbation of $W$, hence the maximal cut in $W'$ (and
in $W'+D'$), is the same as in $W$. In other words, $mc^*$ is a
vector that minimizes the expression:
\[\min_{x \in \{-1,1\}^n} x(W'+D')x.\]
Also, the vector $u$ minimizes the expression
\[\min_{y \in \R^n} (\sum_{i,j} W_{i,j} y_i y_j + \sum_i D_{i,i} y_i^2)/
||y||^2.\] Think of $u$ as being revealed in two steps. First, the
absolute value of each coordinate is revealed, and then, in the
second step, its sign. Thus, in the second step we are looking for
a sign vector $x$ that minimizes the expression:
\[(\sum_{i,j} W_{i,j} \cdot |u_i| x_i \cdot |u_j| x_j +
\sum_i D_{i,i} u_i^2)/||u||^2.\] Clearly, $mc^*$ is such a vector.
Since the input is stable, the optimal cut is unique, and so
$mc^*$ and $-mc^*$ are the only such vectors. Hence, the partition
they induce is the same as that induced by $u$. \QED
\begin{note}
A more careful analysis shows a somewhat stronger result. It
suffices that
\[\gamma \geq \frac {\max_{(i,j) \in E \;:\; u_i u_j < 0} -u_i u_j}
 {\min_{(i,j) \in E \;:\; u_i u_j \geq 0} u_i u_j}.\]
\end{note}

\subsection{A sufficient condition for extended spectral partitioning}

\begin{lemma}\label{lem-sdp}
Let $W$ be a $\gamma$-stable instance of Max-Cut,
for $\gamma > 1$, and let $D$ be the diagonal
matrix $D_{i,i} = mc^*_i \sum_j W_{i,j} mc^*_j$. If $W+D$ is
positive semi-definite, then extended spectral partitioning solves
Max-Cut for $W$ efficiently.
\end{lemma}
\proof It is easy to see that the vector $mc^*$ is in the kernel
of $W+D$. Since $W+D$ is positive semidefinite, $0$ is its least
eigenvalue, and $mc^*$ is an eigenvector of $W+D$ corresponding to
the smallest eigenvalue. Hence, the assertion of Lemma
\ref{lem-spec-part} holds. It remains to show that Max-Cut can
be found efficiently.

Observe that $trace(D) = w(E_{cut}) - w(E_{notcut}) = 2 \cdot
w(E_{cut}) - w(E)$, where $E_{cut}$ is the set of edges in the
maximal cut, and $E_{notcut}$ is the set of all other edges.
Hence, to determine the value of the Max-Cut, it suffices to
compute $m=trace(D)$. Since $mc^*(W+D)mc^* = 0$, it follows that
$mc^*\; W\; mc^* = -m$.

We claim that $m = \min \mbox{~trace}(A)$ over $A \in \A$, where
$\A$ is the set of all positive definite matrices $A$ such that
$A_{i,j} = W_{i,j}$ for $i \neq j$.  (As we discuss in subsection
\ref{sec:gw} below, this is the dual problem of the
Goemans-Williamson relaxation (\cite{GoWi}).)

That the smallest such trace is $\le m$ follows since $W+D \in
\A$. For the reverse inequality note that every $A \in \A$
satisfies $mc^* A mc^* = -m + trace(A)$. But $A$ is positive
semidefinite so $trace(A) \geq m$ as claimed.

As observed by Delorme and Poljak \cite{DelPol} (and, in fact
already in \cite{Bop}), the theory developed by Gr\"{o}tschel,
Lov\'{a}sz and Schrijver \cite{GLS81, GLS84} around the ellipsoid
algorithm makes it possible to efficiently solve the above
optimization problem. 

Note that the solution to the optimization problem is not
necessarily unique, but this can be overcome by slightly
perturbing $W$ at random. If $W$ is stable, then such
a modification leaves $mc^*$ unchanged.\QED

If $W$ is a real symmetric matrix under consideration, we denote
its eigenvalues by $\lambda_1 \ge \dots \ge \lambda_n$. We show
next that if the last two eigenvalues are sufficiently
small in absolute value, then the assertion in Lemma \ref{lem-sdp}
holds. We also recall
the notation $w(i) = \sum_j W_{i,j}$. Since $w(i)$ can be viewed
as a ``weighted vertex degree'', we denote $\min_i\{w(i)\}$
by $\tilde{\delta} = \tilde{\delta}(W)$.

\begin{lemma}\label{lem-spec-part-2}
Let $W$ be a $\gamma$-locally stable instance of Max-Cut with
spectrum $\lambda_1 \ge \dots \ge \lambda_n$, support $G$ and
smallest weighted degree $\tilde{\delta}$.
Let $D$ be a diagonal matrix with $D_{i,i} = mc^*_i
\sum_j W_{i,j} mc^*_j$. If
\[ 2 \tilde{\delta} \cdot \frac {\gamma - 1} {\gamma + 1} + \lambda_n + \lambda_{n-1} > 0, \]
then $W+D$ is positive semidefinite. Furthermore, if $W$ is $\gamma$ stable
for  $\gamma > 1$ then Max-Cut can be found efficiently.
\end{lemma}

\proof
Let $x$ (resp. $y$) be a unit eigenvectors of $W+D$ corresponding
to the smallest (second smallest) eigenvalue of $W+D$. We can
and will assume that $x$ and $y$ are orthogonal. Since
$0$ is an eigenvalue of $W+D$ (with eigenvector $mc^*$) it follows
that $x(W+D)x \leq 0$. If we can show that $y(W+D)y > 0$,
then the second smallest eigenvalue of $W+D$ is
positive, and this matrix is positive semidefinite, as claimed. 

By local stability,
$D_{i,i} \geq \frac {\gamma - 1} {\gamma + 1} \tilde{\delta}$,
so all of $D$'s eigenvalues are at least
$\frac {\gamma - 1} {\gamma + 1} \tilde{\delta}$.

Therefore
\[xWx \leq -xDx \leq -\frac {\gamma - 1} {\gamma + 1} \tilde{\delta}.\]

By the variational theory of eigenvalues (the Courant-Fischer
Theorem), since $x$ and $y$ are two orthogonal unit vectors
there holds

$$
\lambda_n + \lambda_{n-1} \le xWx + yWy.
$$

Also,

$$
\frac {\gamma - 1} {\gamma + 1} \tilde{\delta} \le yDy.
$$

When we sum the three inequalities it follows that

$$
2 \frac {\gamma - 1} {\gamma + 1} \tilde{\delta} + \lambda_n + \lambda_{n-1} \le y(W+D)y.
$$

The Lemma follows. Lemma \ref{lem-sdp}
implies that extended spectral partitioning solves Max-Cut for $W$.
\QED

\subsection{Examples of graph families on which Max-Cut can be found efficiently}
Lemma \ref{lem-spec-part-2} gives a sufficient conditon under
which the extended spectral partitioning solves Max-Cut
efficiently. In this subsection we identify certain families of graphs
for which the assertion in the lemma holds.

\begin{example}\label{example:regular}
Let $G$ be a $1+\epsilon$ stable, $\gamma$-locally stable graph with all $w(i)$ equal.
Let $\lambda_{n-1} \ge \lambda_n$ be its two smallest eigenvalues.
Max-Cut can be found efficiently on $G$ if 
\[\frac {\lambda_{n-1}} {\lambda_n} < \frac {\gamma - 3} {\gamma + 1},\]
and $\epsilon > 0$.
\end{example}
\proof
By the Perron-Frobenius theorem, $\lambda_1 = \tilde{\delta}$, and
the all-one vector is the corresponding eigenvector. It also implies
that $\tilde{\delta} = \lambda_1 \ge |\lambda_n|$.
For the condition in lemma \ref{lem-spec-part-2} to hold, it thus suffices that 
\[ -2 \cdot \lambda_n \frac {\gamma - 1} {\gamma + 1} + \lambda_n + \lambda_{n-1} > 0,\]
which is exactly the stated condition.
\QED

\begin{example}\label{example:expander}
Let $G$ be a $1+\epsilon$ stable, $\gamma$-locally stable $d$-regular simple
graph with second eigenvalue $\lambda$.
Max-Cut can be found efficiently on $G$ if
\[\gamma > \frac{5d + \lambda}{d - \lambda},\]
and $\epsilon > 0$.
\end{example}
\proof
Let $A$ be the adjacency matrix of $G$, and $A_{in}$
the adjacency matrix of the graph spanned by the edges of the maximal cut.
Let $A_{out} = A - A_{in}$.
Since $G$ is $\gamma$-locally stable the maximal degree in $A_{out}$,
and hence its spectral radius, is at most $\frac d {\gamma + 1}$. Therefore,
by subtracting $A_{out}$ from $A$, eigenvalues are shifted by at most this value
(this follows, e.g., by Weyl's theorems on matrix spectra).
In other words, the second eigenvalue of $A_{in}$ is at most $\lambda + \frac d {\gamma + 1}$.
Since $A_{in}$ is bipartite, its spectrum is symmetric, and so
$|\lambda_{n-1}(A_{in})| \leq \lambda + \frac d {\gamma + 1}$.
Now adding  $A_{out}$ to $A_{in}$ again shifts the spectrum by at most  $\frac d {\gamma + 1}$,
and so $|\lambda_{n-1}(A)| \leq \lambda + \frac {2d} {\gamma + 1}$.
In addition, by the Perron-Forbenius theorem,  $|\lambda_n(A)| \leq d$ and so
\[-(\lambda_n(A) + \lambda_{n-1}(A)) \leq d + \lambda + \frac {2d} {\gamma + 1}.\]
For the condition in  lemma \ref{lem-spec-part-2} to hold, it thus suffices that
\[ 2 d \cdot \frac {\gamma - 1} {\gamma + 1} > d + \lambda + \frac {2d} {\gamma + 1}, \]
as claimed.\QED

\begin{example}\label{example:cheeger}
Let $G=(V,E)$ be a $1+\epsilon$ stable, $d$-regular simple graph
with Chegger constant $h$.
Max-Cut can be found efficiently on $G$ if
\[\gamma > \frac {5 + \sqrt{1-(h/d)^2}} {1 - \sqrt{1-(h/d)^2}},\]
and $\epsilon > 0$.
\end{example}
\proof
Recall that the Cheeger constant of a graph is defined as
\[h(G) = \min_{U \subset V \;:\; |U| \leq \frac n 2}
\frac {|E(U,\bar U)|} {|U|},\]
and provides on upper bound on  $G$'s second eigenvalue (e.g. \cite{Mohar}):
\[\lambda_2(G) \leq \sqrt{d^2-h(G)^2}.\]
By Example~\ref{example:expander} Max-Cut can be found efficiently on $G$.
\QED

\begin{example}\label{example:distinct}
Let $G=(V,E)$ be a $1+\epsilon$ stable, $(k,\gamma)$-distinct $d$-regular simple graph.
Max-Cut can be found efficiently on $G$ if
\[\gamma > \frac {5 + \sqrt{1-(k/d)^2}} {1 - \sqrt{1-(k/d)^2}},\]
and $\epsilon > 0$.
\end{example}
\proof
Let $(S, \bar S)$ be the largest cut in $G$. Pick an arbitrary set
$U \subset V$ of size $\le n/2$. We will derive a lower bound on
$|E(U, \bar U)|$ and therefore a lower bound on $G$'s Cheeger constant.

So let us consider the cut $(T, \bar T)$
obtained from $(S, \bar S)$ by swapping the position of each vertex in $U$.
Since $|U| < n/2$,
\[\min\{|S \Delta T|, |S \Delta  \bar T|\} = \min\{ |U|, |\bar U| \} = |U|.\]
Now $k$-distinctness implies that $|E(T, \bar T)| \le |E(S, \bar S)| - k|U|$.
But every edge in $E(S, \bar S) \setminus E(T, \bar T)$ belongs to
$E(U, \bar U)$. Consequently, $|E(U, \bar U)| \geq k|U|$, and since $U$
was arbitrary, $h \geq k$.

By Example~\ref{example:cheeger} Max-Cut can be found efficiently on $G$.
\QED

\section{Results derived from previous works}
\subsection{Performance of the Goemans-Williamson approximation algorithm}
\label{sec:gw}
Let us quickly recall the Goemans and Williamson approximation
algorithm for Max-Cut~\cite{GoWi}. We first rephrase the Max-Cut problem as:
\begin{eqnarray*}
&&\mbox{Maximize } \half \sum_{(i,j) \in E}W_{i,j}(1-y_i y_j)\\
&&\mbox{over } y \in \{-1,1\}^n.
\end{eqnarray*}
Equivalently, we seek to minimize $ \sum_{(i,j) \in E} W_{i,j} Y_{i,j}$
over all $\{-1,1\}$-matrices
$Y$ that are positive semi-definite and of rank 1. In the G-W algorithm
the rank constraint is relaxed, yielding a semi-definite programming problem
which can be solved efficiently with approximation guarantee of $\sim 0.8786$.
Moreover, they show that when the weight
of the maximal cut is sufficiently big, this guarantee can be improved.
Namely, let $R \; (\geq \half)$ be the ratio between the weight of the
maximal cut and the
total weight of the edges. Let $h(t) = arccos(1-2t)/\pi$. Then the
approximation ratio is at least $h(R)/R$.\\
By local stability, the contribution of each $v \in V$ to the
maximal
cut is $\frac {\gamma} {\gamma + 1}$ the total weight of the edges incident with it.
Summing this over all vertices, we get that the maximal cut weighs at least
$R = \frac {\gamma} {\gamma + 1}$ of the total weight.
Thus, the performance guarantee of the G-W algorithm on $\gamma$-stable
instances is at least $(1 - O(\frac 1 {\sqrt{\gamma}}))$.

Note that for this we only required local stability.

The semi-definite program used in the G-W algorithm can be
strengthened when the input is
$\gamma$-stable, by inequalities that express this stability.
It is interesting whether these additional constraints can improve the
approximation ratio further.

\subsection{Spectrally partitioning random graphs}
Consider the following model for random weighted graphs.
Let $P$ be some probability measure on $[0,\infty)$. Generate
a matrix $W'$ (a weighted adjacency matrix), by choosing each entry $W'_{i,j}$,
$i < j$, independently from $P$. Set $W'_{i,j} = W'_{j,i}$ for $ i > j$,
and $W'_{i,i} = 0$. Let $C$ be the set of edges in the maximal cut of $W$
(for ``reasonable'' $P$'s, this will be unique w.h.p.). Set
$W_{i,j} = \gamma \cdot W'_{i,j}$ for $(i,j) \in C$.\\
It is easy to see that $W$ is indeed $\gamma$-stable, yet for certain
probability measures the problem becomes trivial. For example, if $P$ is
a distribution on $\{0,1\}$, the maximal cut in $W$ simply
consists of all the edges
with weight $\gamma$.\\
An even simpler random model is the following. Take $n$ even. Generate
an $n \times n$ matrix
$W'$ as above. Choose $S \subset [n]$, $|S| = n/2$ uniformly at random.
Let $C$ be the set of edges in the cut $(S,\bar{S})$.
Set $W_{i,j} = \gamma \cdot W'_{i,j}$ for $(i,j) \in C$. Denote this
distribution $\G(n,P,\gamma)$.
For an appropriate
$\gamma$, w.h.p. $(S,\bar{S})$ will be the maximal cut in $W$. This random
model is close to what is sometimes known as ``the planted partition model''
(\cite{Bui, Bop, DyFr, JeSo, CoKa, FeKi, McSherry, ShTs}).\\
Following work by Boppana \cite{Bop} on a similar random model (for unweighted graphs),
we can deduce that w.h.p. the maximal cut of graphs from this
distribution can be found efficiently:
\begin{theorem}
Let $P$ be a distribution with bounded support, expectation $\mu$ and variance
$\sigma^2$. There exists a polynomial time algorithm that w.h.p. solves Max-Cut
for $G \in \G(n,P,\gamma)$, when $\gamma = 1 + \Omega(\sqrt{\frac {\log n} n})$.
\end{theorem}

The theorem follows from
Lemma \ref{lem-sdp} and the following one, which is an easy consequence of \cite{Bop}:
\begin{lemma}\label{lem-rnd-grph}
Let $P$ be a distribution with bounded support, expectation $\mu$ and variance
$\sigma^2$. Let $G \in \G(n,P,\gamma)$, and $S$ the subset
chosen in the generating $G$. Let $mc \in \{-1,1\}^{n}$ be the indicator
vector of the cut $(S,\bar{S})$.
Let $D$ be the diagonal matrix defined
by $D_{i,i} = mc \; W \; mc$. If
$\gamma \geq 1 + \Omega(\sqrt{\frac {\log n} n})$, then w.h.p.:\\
1. $mc$ is the indicator vector of the maximal cut in $G$.\\
2. $W+D$ is positive semi-definite.
\end{lemma}

\section{Conclusion and open problems}
In this work we have shown that stability, supplemented by certain properties
of the input instance, allows for an efficient algorithm for Max-Cut.
However, if nothing is assumed about the input, we only know that
$n$-stability is sufficient.
Can this be improved? Note that $\gamma \geq n$ is very far from what happens in the
random model, where it is only required that $\gamma \geq 1 + \Omega(\sqrt{\frac {\log n} n})$.
A bold conjecture
is that there is some constant, $\gamma^*$, s.t. $\gamma^*$-stable instances
can be solved in polynomial time.

Our motivation in defining stability and distinctness is to
identify natural properties
of a solution to an NP-hard problem, which ``make it interesting'',
and allow finding it in
polynomial time. Stability and distinctness
indeed make Max-Cut amenable, but are in no way the only possible properties, and
it would be very interesting to suggest others.

\bibliography{bib}

\begin{thebibliography}{10}

\bibitem{Bop}
R.~Boppana.
\newblock Eigenvalues and graph bisection: An average case analysis.
\newblock In {\em 28th Annual Symposium on Foundations of Computer Science,
  October 12--14, 1987, Los Angeles, California}, pages 280--285. IEEE Comput.
  Soc. Press, 1987.

\bibitem{Bui}
T.~N. Bui, S.~Chaudhuri, F.~T. Leighton, and M.~Sipser.
\newblock Graph bisection algorithms with good average case behavior.
\newblock {\em Combinatorica}, 7(2):171--191, 1987.

\bibitem{CoKa}
A.~Condon and R.~M. Karp.
\newblock Algorithms for graph partitioning on the planted partition model.
\newblock {\em Random Structures Algorithms}, 18(2):116--140, 2001.

\bibitem{DelPol}
C.~Delorme and S.~Poljak.
\newblock Laplacian eigenvalues and the maximum cut problem.
\newblock {\em Math. Programming}, 62(3, Ser. A):557--574, 1993.

\bibitem{DyFr}
M.~E. Dyer and A.~Frieze.
\newblock Fast solution of some random np-hard problems.
\newblock In {\em 27th Annual Symposium on Foundations of Computer Science,
  October 27--29, 1986, Toronto, Ontario, Canada}, pages 313--321. IEEE Comput.
  Soc. Press, 1986.

\bibitem{FeKi}
U.~Feige and J.~Kilian.
\newblock Heuristics for semirandom graph problems.
\newblock {\em J. Comput. System Sci.}, 63(4):639--671, 2001.
\newblock Special issue on FOCS 98 (Palo Alto, CA).

\bibitem{GoWi}
M.~X. Goemans and D.~P. Williamson.
\newblock Improved approximation algorithms for maximum cut and satisfiability
  problems using semidefinite programming.
\newblock {\em J. Assoc. Comput. Mach.}, 42(6):1115--1145, 1995.

\bibitem{GLS81}
M.~Gr{\"o}tschel, L.~Lov{\'a}sz, and A.~Schrijver.
\newblock The ellipsoid method and its consequences in combinatorial
  optimization.
\newblock {\em Combinatorica}, 1(2):169--197, 1981.

\bibitem{GLS84}
M.~Gr{\"o}tschel, L.~Lov{\'a}sz, and A.~Schrijver.
\newblock Corrigendum to our paper: ``{T}he ellipsoid method and its
  consequences in combinatorial optimization'' [{C}ombinatorica {\bf 1} (1981),
  no. 2, 169--197.
\newblock {\em Combinatorica}, 4(4):291--295, 1984.

\bibitem{JeSo}
M.~Jerrum and G.~B. Sorkin.
\newblock The {M}etropolis algorithm for graph bisection.
\newblock {\em Discrete Appl. Math.}, 82(1-3):155--175, 1998.

\bibitem{McSherry}
F.~McSherry.
\newblock Spectral partitioning of random graphs.
\newblock In {\em 42nd IEEE Symposium on Foundations of Computer Science (Las
  Vegas, NV, 2001)}, pages 529--537. IEEE Computer Soc., Los Alamitos, CA,
  2001.

\bibitem{Mohar}
B.~Mohar.
\newblock Isoperimetric numbers of graphs.
\newblock {\em J. Combin. Theory Ser. B}, 291:47--274, 1989.

\bibitem{Papadi}
C.~H. Papadimitriou.
\newblock {\em Computational complexity}.
\newblock Addison-Wesley Publishing Company, Reading, MA, 1994.

\bibitem{ShTs}
R.~Shamir and D.~Tsur.
\newblock Improved algorithms for the random cluster graph model.
\newblock In {\em Proceedings of the 8th Scandinavian Workshop on Algorithm
  Theory}, pages 230--239. 2002.

\bibitem{SpTeng}
D.~Spielman and S.~H. Teng.
\newblock Smoothed analysis of algorithms: why the simplex algorithm usually
  takes polynomial time.
\newblock In {ACM}, editor, {\em Proceedings of the 33rd Annual ACM Symposium
  on Theory of Computing: Hersonissos, Crete, Greece, July 6--8, 2001}, pages
  296--305, New York, NY, USA, 2001. ACM Press.
\newblock ACM order number 508010.

\bibitem{Vershy}
R.~Vershynin.
\newblock Beyond hirsch conjecture: Walks on random polytopes and smoothed
  complexity of the simplex method.
\newblock In {\em FOCS}, pages 133--142. IEEE Computer Society, 2006.

\end{thebibliography}

\end{document}